\documentclass[12pt]{article}

\parskip=\medskipamount                         
\def\7#1#2{\mathop{\null#2}\limits^{#1}}        

\def\beee{\begin{equation}}
\def\eeee{\end{equation}}

\usepackage{graphicx}
\usepackage{setspace}
\usepackage{amssymb}
\usepackage[T1]{fontenc}
\usepackage[latin9]{inputenc}
\usepackage{color}
\usepackage{esint}
\usepackage{epsf}
\begin{document}
\begin{center}
 \textbf { \large A self-consistent bound state model for meson}\\
[5mm] Asrarul Haque\footnote{email address: ahaque@iitk.ac.in}\\
{ Department of Physics, I.I.T. Kanpur, Kanpur 208016(INDIA)}\\
Satish D. Joglekar\footnote{email address: sdj@iitk.ac.in}\\
{ Department of Physics, I.I.T. Kanpur, Kanpur 208016(INDIA)}\\
and\\
NISER, Bhubaneswar 751005 (INDIA)
\end{center}
\begin{abstract}
We study the 1+1 dimensional Yukawa theory, in a certain limit of its parameters
$g,M,m$ ( as suggested by the study of causality in presence of bound states in
this model \cite{C}). We study the bound state formation in the model. In the limit
$g\rightarrow\infty,M\rightarrow\infty$, in a certain specific manner, we show that
there are a large number of bound states of which at least the low lying states
are described by the non-relativistic Schrodinger equation. We show that, in this
limit,  the excited bound states are unstable and deem to decay quickly (lifetime
$\tau\rightarrow0$) by emission of scalar (s) in this particular limit. The mass of
the ground state is not significantly affected by higher order quantum corrections
and by proper choice of parameters, involving only small changes, can be
adjusted to be equal to the mass of the scalar. As a result of quantum effects, the
state of the meson mixes with the lowest bound state and may be dominated by
the latter .We show that in this \emph{detailed sense}, a scalar meson in Yukawa
model can be looked upon as a bound state of a fermion-anti fermion pair formed.
\end{abstract}
\section{Preliminary}
 We start with the $1+1$ dimensional Yukawa
theory,\begin{equation}
\mathcal{L}=\bar{\psi}\left[i\not{\partial}
-M+g\phi\right]\psi+\frac{1}{2}\partial_{\mu}\phi\partial^{\mu}\phi-\frac{1}{2}m^{2}\phi^{2}\label{eq:lag}\end{equation}
We expand $\phi=\phi_{c}+\phi_{q}$, where $\phi_{c}$ is the
classical field obeying the classical equation of motion:
\begin{equation}
\left[\partial^{2}+m^{2}\right]\phi_{c}=g\bar{\langle{\psi}}\psi\rangle,\label{eq:CY}\end{equation}
leading to the Yukawa potential, and \begin{equation}
\mathcal{L_{I}=}g\bar{\psi}\psi\phi_{q}+.....\label{eq:int}\end{equation} is the
quantum interaction Lagrangian. {[}This division is more or less like what we do
when we discuss radiation from an atom. (see e.g. \cite{MB})] The fermion
anti-fermion pair interacts in the first instance via $\phi_{c}$, the classical
Yukawa potential and form bound states. In absence of $\mathcal{L_{I}}$, these
are the exact \emph{stationary} states of the system (like the H-atom has them if
interaction of an electron with a photon were neglected). The excited states,
however, are unstable against decay via emission of scalar(s) through the
interaction Lagrangian $\mathcal{L_{I}}$. We want to show that in a certain limit
of parameters $g,m,M$, (suggested by a study of causality in this model,
\cite{C}) there is only one \emph{stable} bound state and the scalar particle can
be identified with it. All other bound states are highly unstable.
($\tau\rightarrow0$).
\section{Non-relativistic Bound States}
We shall consider the fermion anti-fermion system (mass $M$ each)
in a bound state via the attractive Yukawa potential {[}This is a
solution of (\ref{eq:CY}), under appropriate conditions].
\[
V\left(x\right)=-g^{2}\frac{e^{-m|x|}}{2m}\]
We shall assume that the mass $M$ is large, and that the bound state
is non-relativistic. (This will be justified later.) For the relative
motion, the non-relativistic Schrodinger equation is satisfied (reduced
mass is $\frac{M}{2}$):\[
-\frac{\hbar^{2}}{M}\frac{d^{2}\psi}{dx^{2}}-g^{2}\frac{e^{-m|x|}}{2m}\psi=E\psi\]
where $E=-\varepsilon<0$ is the energy of a bound state. We introduce
the change of the independent variable: $y=e^{-m|x|/2}$. Under this
change, the equation transforms to
\[
y^{2}\frac{d^{2}\psi}{dy^{2}}+y\frac{d\psi}{dy}+\frac{2Mg^{2}}{m^{3}}y^{2}\psi+\frac{4EM}{m^{2}}\psi=0\]
{[}The equation, now, has to be solved separately for $x>0$ and $x<0$
and the solutions have to be tested for continuity of $\psi$ and
$\psi'$ at $x=0$]. A further rescaling: $t=\xi y\equiv\sqrt{\frac{2Mg^{2}}{m^{3}}}y=\sqrt{\frac{2Mg^{2}}{m^{3}}}e^{-m|x|/2}$
will put the equation in the form of the Bessel equation \cite{G,HPS}:\[
\psi''+\frac{1}{t}\psi'+\left(1-\frac{\nu^{2}}{t^{2}}\right)\psi=0.\]
with $\nu\equiv+\sqrt{\frac{4\varepsilon M}{m^{2}}}$. {[}We recall
that as $\varepsilon$ is always less than the depth of the well $\frac{g^{2}}{2m}$,
the allowed values of $\nu$ will turn out to be less than $\xi$].
We note that as $x\rightarrow$$\pm\infty,\: t\rightarrow0.$ As $x\rightarrow0^{\pm},\: t\rightarrow\xi$.
The boundary condition at $x\rightarrow\pm\infty$ requires that $\psi\left(0\right)\rightarrow0$.
This requires that we take the solution which is regular at $t=0$,
viz. $J_{\nu}\left(t\right)$ and drop the singular solution $N_{\nu}\left(t\right)$.
The discrete values of $\varepsilon$ are determined from the conditions
of continuity at $x=0$:
\begin{enumerate}
\item $\psi\left(x\rightarrow0^{-}\right)=\psi\left(x\rightarrow0^{+}\right)\Rightarrow\psi\left(t\right)$
is continuous as $t\rightarrow\xi$ from either side.
\item $\frac{d}{dx}\psi\left(x\rightarrow0^{-}\right)=\frac{d}{dx}\psi\left(x\rightarrow0^{+}\right)\Rightarrow\psi'\left(t\rightarrow\xi;\: x\rightarrow0^{-}\right)=-\psi'\left(t\rightarrow\xi;\: x\rightarrow0^{+}\right)$
\end{enumerate}
These boundary conditions can materialize in two distinct ways: As
the potential is even, the non-degenerate states will have a definite
parity\cite{LL}; so that,
\begin{enumerate}
\item the wave-function is even and $\frac{d}{dx}\psi\left(x\rightarrow0^{-}\right)=\frac{d}{dx}\psi\left(x\rightarrow0^{+}\right)=0$;
(and $\psi\left(x=0\right)\neq0$) ;
\item the wave-function is odd: $\psi\left(x=0\right)=0,$ (and $\frac{d}{dx}\psi\left(x\rightarrow0^{-}\right)=\frac{d}{dx}\psi\left(x\rightarrow0^{+}\right)\neq0$)
;
\end{enumerate}
In the case 1 above, $\psi\left(t\right)$ satisfies: $\psi'\left(t\right)=0\: at\: t=\xi$.
In the case 2 above, $\psi\left(t\right)$ satisfies: $\psi\left(t\right)=0\: at\: t=\xi$.
Thus, the energy eigenvalues are determined by:
\begin{enumerate}
\item $J_{\nu}\left(\xi\right)=0$, or
\item $J'_{\nu}\left(\xi\right)=0$
\end{enumerate}
{[}where, from the properties of $J_{\nu}\left(\xi\right)$, both
$J_{\nu}\left(\xi\right)$ and $J'_{\nu}\left(\xi\right)$ cannot
simultaneously vanish for a $\xi>0$ \cite{W} ]. The energy dependence
enters through the order $\nu\equiv+\sqrt{\frac{4\varepsilon M}{m^{2}}}$
of $J_{\nu}$.
\section{Properties of Bound States and Relevant Properties of $J_{\nu}\left(\xi\right)$}
For the ground state, being  an even wavefunction,,
$J'_{\nu}\left(\xi\right)=0$. $\nu$ is so determined that $\xi$ is
the first positive zero of $J'_{\nu}\left(\xi\right)$. The ground
state wave-function is \begin{equation}
\psi=CJ_{\nu}\left(\xi\exp\left[-m|x|/2\right]\right)\label{eq:WF}\end{equation}
The relevant properties of the ground state wave-function are:
\begin{enumerate}
\item $\nu<\xi$ as expected and for a large $\xi$, $\frac{\nu}{\xi}\rightarrow1$
from below. This implies that $\frac{\varepsilon}{g^{2}/2m}\rightarrow1$
from below. The exact behavior is given as \cite{W1},\begin{equation}
\xi=\nu+0.808618\nu^{1/3}+O\left(\nu^{-1/3}\right)\equiv\nu+\alpha\nu^{1/3}+O\left(\nu^{-1/3}\right)\label{eq:XINU}\end{equation}
This can be numerically verified: See Table 1.
\item $\frac{\xi^{2}-\nu^{2}}{\xi^{2}}\rightarrow0$. The quantity $\xi^{2}$
is proportional to the maximum depth of the well $\frac{g^{2}}{2m}$,
$\nu^{2}$ like-wise is proportional to the magnitude of the ground
state energy eigenvalue. So, $\left(\xi^{2}-\nu^{2}\right)$ gives
an upper bound on the kinetic energy of the ground state. We shall
consider the depth of the potential approximately equal to $2M$.
The relation $\frac{\xi^{2}-\nu^{2}}{\xi^{2}}\rightarrow0$ then implies:
$\frac{<KE>}{g^{2}/2m}=\frac{<KE>}{2M}\rightarrow0$ ; or that the
ground state is non-relativistic. Evidently, it will also apply to
low lying excited states.
\item $\frac{\xi^{2}-\nu^{2}}{\xi^{2}}\rightarrow0$. It behaves as $\frac{\xi^{2}-\nu^{2}}{\xi^{2}}\sim\nu^{-2/3}$.
This follows from (\ref{eq:XINU}), which implies,\begin{eqnarray}
\left(\xi^{2}-\nu^{2}\right) & = & \left(\xi-\nu\right)\left(\xi+\nu\right)\nonumber \\
 & = & \left[0.808618\nu^{1/3}+O\left(\nu^{-1/3}\right)\right]\nonumber \\
 & \,\times & \left[2\nu+0.808618\nu^{1/3}+O\left(\nu^{-1/3}\right)\right]\\
 & = & 1.617236\nu^{4/3}+0.654\nu^{2/3}+O\left(1\right)\nonumber \\
\frac{\left(\xi^{2}-\nu^{2}\right)}{\xi^{2}}\nu^{2/3} & = & 1.617236\frac{\nu^{2}}{\xi^{2}}+O\left(\nu^{-2/3}\right)\label{eq:ke/M}\\
 & = & 1.617236+O\left(\nu^{-2/3}\right)\nonumber \end{eqnarray}
We first note that the (\ref{eq:ke/M}) tells us that \begin{equation}
\frac{KE}{M}\sim\nu^{-2/3}<<1\label{eq:keM}\end{equation}
and offers a justification for treating the ground state non-relativistically.
Further, the specific behavior implies that the root mean square
momentum $\sqrt{\left\langle p^{2}\right\rangle }\lesssim M^{5/9}$.
This behavior will be used in deciding the cut-off $\Lambda$ on the
quantum correction to propagator.
\item Let $\nu'$ be the largest value  satisfying $J_{\nu'}\left(\xi\right)=0$
for a given $\xi$ . This state corresponds to the first exited state.
Then \cite{W2},\begin{equation}
\xi\approx\nu'+1.8588\nu'^{1/3}\equiv\nu'+\beta\nu'^{1/3}\label{eq:XINU'}\end{equation}
Equations (\ref{eq:XINU}) and (\ref{eq:XINU'}) imply \[
\frac{\nu^{2}-\nu'^{2}}{\nu^{2}}=2\left(\beta-\alpha\right)\nu^{-2/3}+O\left(\nu^{-4/3}\right)=2.1\nu^{-2/3}+O\left(\nu^{-4/3}\right)\]
Recalling that $\nu^{2}-\nu'^{2}\propto\Delta\varepsilon$, the energy
gap between the first excited state and the ground state. This implies
that,\begin{eqnarray*}
\frac{\Delta\varepsilon}{2M} & = & 2.1\left(\frac{m}{2\sqrt{2}M}\right)^{2/3}\\
\frac{\Delta\varepsilon}{m} & = & 2.1\left(\frac{2M}{m}\right)\left(\frac{m}{2\sqrt{2}M}\right)^{2/3}=2.1\times\left(\frac{M}{m}\right)^{1/3}>1\end{eqnarray*}
Thus, the energy gap $\Delta\varepsilon>>m$. Kinematics does not
forbid excited state be unstable against the decay:\[
Excited\; state\rightarrow ground\; state+a\; scalar\]
This holds also for higher  excited states.
\item The excited states can decay to the ground state or lower excited states by emission of a scalar(s).
We shall present the calculation regarding this in section \ref{sec:decay}.
The decay width is $\sim g^{2}$ and is large as $g\;\mbox{and}\;M \rightarrow\infty$.
The excited states are very unstable in this limit. The life-time
of excited states will $\rightarrow 0$ as $g$ increases.
\end{enumerate}
These set of results above can be verified numerically \cite{BC}
(See table 1): In table 1, we have numerically evaluated various
quantities. We found it convenient  to take a several simple
values of $\nu$ and found the smallest solution $\xi$ for it. We
see that consistently, the ratio $\frac{\nu}{\xi}$ approaches 1
from below. We next study how $\frac{\nu}{\xi}$ approaches 1 from
below. We compute $\frac{\xi^{2}-\nu^{2}}{\xi^{2}}\nu^{2/3}$ and
note that it approaches a constant. For each $\xi$, we find the
largest solution $\nu'$ satisfying $J_{\nu'}\left(\xi\right)=0$
corresponding to the first excited state. We have had to round the
value to an integer (which explain the fluctuations in the last
column.)
\begin{table}
\begin{center}
Table 1\vspace{0.5cm}
\begin{tabular}{|c|c|c|c|c|c|c|c|c|}
\hline $\nu$ & $\xi$ & $\frac{\nu}{\xi}$
&$\frac{\xi^2-\nu^2}{\xi^2}$&
$\frac{\xi^2-\nu^2}{\xi^2}\nu^{\frac{2}{3}}$
&$\nu'$&$\nu^2-\nu'^2$&$\frac{\nu^2-\nu'^2}{\nu^{1.333}}$\tabularnewline
\hline \hline 100 & 103.76838 & 0.9637 & 0.0713118& 1.535893693 &
95&975&2.1038008 \tabularnewline \hline 150 & 154.30972 & 0.9721 &
0.055078& 1.554393383 & 144&1764&2.21701654\tabularnewline \hline
199 & 203.73309 & 0.9768 & 0.0459239& 1.564782456 &
193&2352&2.02799167\tabularnewline \hline 300 & 305.4238 & 0.9822
& 0.0352012& 1.576909638 & 283&4151&2.07085695\tabularnewline
\hline 500 & 506.42703 & 0.9873 & 0.0252208& 1.588151557 &
492&7936&2.00389353\tabularnewline \hline
\end{tabular}
 \end{center}
\end{table}
\section{Connecting Mass with Net Bound State Energy}
We have taken a fermion-anti-fermion pair of mass $M$ each, bound
together in a Yukawa potential of maximum depth
$\approx\frac{g^{2}}{2m}$. We have assumed that $M>>m$. We want to
ultimately show that the bound state can actually be identified
with the scalar mass $m$ of the particle that gives rise to the
Yukawa potential. In order that this can be done, first of all,
there is a need for consistency in the mass of the bound state and
of the scalar. We shall show that by making a small adjustment
(much much less than M) in the depth of the well
$\frac{g^{2}}{2m}$ we can make the net energy equal the mass of
the scalar $m$. We would like, \begin{eqnarray*}
&&\textup{Total relativistic energy of a bound state}  =  2M-\varepsilon=m\\
&&i.e.\;\;2M-\frac{m^{2}\nu^{2}}{4M}  =  m\\
&&\frac{m^{2}\nu^{2}}{8M^{2}}  =
1-\frac{m}{2M}\approx1-\frac{\sqrt{2}}{\nu}\end{eqnarray*} This
together with (\ref{eq:XINU}) leads to\[
\frac{\frac{g^{2}}{2m}}{2M}=\xi^{2}\frac{m^{2}}{8M^{2}}=\left(\frac{\xi}{\nu}\right)^{2}\left(1-\frac{m}{2M}\right)\approx\left(1+\alpha\nu^{-2/3}\right)^{2}\left(1-\frac{\sqrt{2}}{\nu}\right)\approx1-\frac{\sqrt{2}}{\nu}+\frac{2\alpha}{\nu^{2/3}}+...\]
$ $\[
\frac{g^{2}/2m}{2M}=1+\frac{\left(2\alpha\nu^{1/3}-\sqrt{2}\right)}{\nu}+O\left(\nu^{-4/3}\right)\]
Compared to $\frac{g^{2}/2m}{2M}=1$, this is a slight relative
adjustment in the depth, which moreover vanishes as
$M\rightarrow\infty$. We shall see the purpose of this minute
adjustment more clearly in a future section.
\section{Decay of Excited States \label{sec:decay}}
We would like to argue next that  the exited states are highly
unstable and decay to suitable lower excited state(s), including
the ground state. The lifetime of the excited states tends to zero
as $g\rightarrow\infty$. [We cannot base the result simply on dimensional analysis because the result depends on an overlap integral involving $J_{\nu}$.${\nu}$ depends upon the dimensionless ratio $M/m$].To begin with, we shall assume that $g$
is not too large, and we can employ time-dependent perturbation
theory in the Schrodinger picture. Let the $n$-th bound state
{[}$\left(n-1\right)-$th excited state] have energy
$2M-\varepsilon_{n}$ when at rest. Let the bound state under
consideration be in the $n^{th}$ state at $t=0$. We write the
Hamiltonian operator in the Schrodinger picture
as,\begin{eqnarray*}
H & = & H_{H}\left(t=0\right)\\
 & = & H_{0}+gH_{I}\left(t=0\right)\\
H_{I}\left(t=0\right) & = & \int
dx:\overline{\psi}\left(x,0\right)\psi\left(x,0\right)\phi_{q}\left(x,0\right):\end{eqnarray*}
Here, $H_{0}$ is the free Hamiltonian, with the classical Yukawa
interaction, that in the non-relativistic approximation leads to
bound states of a fermion anti-fermion pair and $H_{I}$ is the
interaction Hamiltonian (from (\ref{eq:int})) that leads in
particular, in first order perturbation theory, to a decay to a
lower state with the emission of a scalar. Let the wave-function
of the $m^{th}$ bound state in the momentum space be
$\beta_{m}\left(q\right)$, where $2q$ is the relative momentum
between the quark anti-quark pair. We denote by $\int
dp\beta_{n}\left(p\right)\left|p,-p,\mathbf{0}\right\rangle
\otimes\left|0\right\rangle _{b}$ the initial state, where $p$ and
$-p$ are the momenta of fermion and anti-fermion relative to its
rest-frame, $\mathbf{0}$ is the momentum of the CM and
$\left|0\right\rangle _{b}$ is the vacuum state for bosons. The
final state is $\int
dq\beta_{m}\left(q\right)\left|\frac{\mathbf{P}}{2}+q,\frac{\mathbf{P}}{2}-q,\mathbf{P}\right\rangle
\otimes\left|l\right\rangle $. We express the relevant terms in
$H_{I}$ in terms of creation and destruction
operators:\begin{eqnarray*}
 &  & H_{I}\left(t=0\right)\\
 & = & \int\frac{dl_{1}dp_{1}dp_{2}}{\left(2\pi\right)^{3}\sqrt{8E_{l_{1}}E_{p_{1}}E_{p_{2}}}}\left\{ a_{l_{1}}^{\dagger}\left[d_{p_{1}}^{\dagger}d_{p_{2}}\overline{u}\left(p_{1}\right)u\left(p_{2}\right)2\pi\delta\left(l_{1}+p_{1}-p_{2}\right)\right]\right.\\
 & - & \left.b_{p_{2}}^{\dagger}b_{p_{1}}\overline{v}\left(p_{1}\right)v\left(p_{2}\right)2\pi\delta\left(l_{1}+p_{2}-p_{1}\right)+....\right\} \\
 & \equiv & H_{I}^{(1)}+H_{I}^{(2)}\end{eqnarray*}
where we have expanded%
\footnote{We shall use the conventions in \cite{PS}.%
},\begin{eqnarray*}
\psi\left(x\right) & = & \int\frac{dp}{\left(2\pi\right)\sqrt{2E_{p}}}\left[d_{p}u\left(p\right)e^{-ipx}+b_{p}^{\dagger}v\left(p\right)e^{ipx}\right]\\
\phi\left(x\right) & = & \int\frac{dp}{\left(2\pi\right)\sqrt{2E_{p}}}\left[a_{p}e^{-ipx}+a_{p}^{\dagger}e^{ipx}\right]\\
\left\{ d\left(p\right),d^{\dagger}\left(q\right)\right\}  & = & 2\pi\delta\left(p-q\right)\\
\left|p\right\rangle  & = & \sqrt{2E_{p}}d^{\dagger}\left(p\right)\left|0\right\rangle \\
\left\langle p\right|\left.p'\right\rangle  & = & 2\pi\delta\left(p-p'\right)2E_{p}\end{eqnarray*}
etc.. $H_{I}^{(1)}$ consists of a term that leave anti-quarks unaffected
but destroy and create a quark and a scalar. $H_{I}^{(2)}$ on the
other hand consists of a term that leave quarks unaffected but destroy
and create a anti-quark and a scalar. We note that $d_{p}\left|p'\right\rangle =2\pi\sqrt{2E_{p}}\delta\left(p-q\right)$.
We have to compute the matrix element,\begin{eqnarray*}
\widetilde{M} & \equiv & \int dpdq\beta_{n}\left(p\right)\beta_{m}^{*}\left(q\right)\left\langle l\right|\otimes\left\langle \frac{\mathbf{P}}{2}+q,\frac{\mathbf{P}}{2}-q,\mathbf{P}\right|H_{I}\left(t=0\right)\left|p,-p,\mathbf{0}\right\rangle \otimes\left|0\right\rangle \\
 & \equiv & \mathcal{\widetilde{M\mathrm{_{1}}}\textrm{+}\widetilde{M_{\mathrm{2}}}}\end{eqnarray*}
where,\begin{eqnarray*}
\mathcal{\widetilde{M_{\mathrm{1}}}} & = & \int dpdq\left\langle \frac{\mathbf{P}}{2}+q,\frac{\mathbf{P}}{2}-q,\mathbf{P}\right|\\
 &  & \times\int\frac{dl_{1}dp_{1}dp_{2}}{\left(2\pi\right)^{2}\sqrt{4E_{p_{1}}E_{p_{2}}}}\left\{ \beta_{n}\left(p\right)\beta_{m}^{*}\left(q\right)\left[d_{p_{1}}^{\dagger}d_{p_{2}}\overline{u}\left(p_{1}\right)u\left(p_{2}\right)\delta\left(l_{1}+p_{1}-p_{2}\right)\right]\right\} \\
 & \times & \delta\left(l_{1}-l\right)\left|p,-p,\mathbf{0}\right\rangle \\
 & = &- \int dpdqdl_{1}dp_{1}dp_{2}\left\{ \left[\overline{u}\left(p_{1}\right)u\left(p_{2}\right)\delta\left(l_{1}+p_{1}-p_{2}\right)\right]\right\} \\
 &  & \times\delta\left(l_{1}-l\right)\delta\left(p_{2}-p\right)\delta\left(p_{1}-q-\frac{P}{2}\right)2\pi\delta\left(-p+q-\frac{P}{2}\right)2E_{p}\beta_{n}\left(p\right)\beta_{m}^{*}\left(q\right)\\
 & = &- \int dq\overline{u}\left(q+\frac{P}{2}\right)u\left(q-\frac{P}{2}\right)\beta_{n}\left(q-\frac{P}{2}\right)\beta_{m}^{*}\left(q\right)2\pi2E_{q-\frac{P}{2}}\delta\left(l+P\right)\\
 & \equiv & -\delta\left(l+P\right)\mathcal{M_{\textrm{1}}}\\
\widetilde{\mathcal{M}_{\mathrm{2}}} & = &- \int dq\overline{v}\left(-q-\frac{P}{2}\right)v\left(-q+\frac{P}{2}\right)\beta_{n}\left(q+\frac{P}{2}\right)\beta_{m}^{*}\left(q\right)\delta\left(l+P\right)2\pi2E_{q+\frac{P}{2}}\\
 & \equiv & -\delta\left(l+P\right)\mathcal{M_{\textrm{2}}}\end{eqnarray*}
Let \[ H_{mn}=\left\langle m\right|gH_{I}\left|n\right\rangle =
-g\delta\left(l+P\right)\mathcal{\left[M_{\textrm{1}}\textrm{+}M_{\textrm{2}}\right]}\equiv
-g\delta\left(l+P\right)\mathcal{M}\] We shall be using formulas
corresponding to discrete normalization rather than continuous.
Hence, we make the replacements \cite{BD}:\[
\frac{L\delta_{p,p'}}{2\pi}\leftrightarrow\delta\left(p-p'\right);\quad
d_{p}^{\dagger}=1/\sqrt{\left(2L\right)}d^{\dagger}\left(p\right);\quad\left\Vert
p\right\rangle =1/\sqrt{2LE_{p}}\left|p\right\rangle ;\quad
a_{p}^{\dagger}=1/\sqrt{\left(2L\right)}a^{\dagger}\left(p\right)\]
{[}Here, $d_{p}^{\dagger}$ and $\left\Vert p\right\rangle $ refers
to the discrete formulation and $d^{\dagger}\left(p\right)$ and
$\left|p\right\rangle $ to continuum formulation.]. We have, for
the continuum matrix element, now called $\widetilde{H}_{mn}$,\[
\widetilde{H}_{mn}\leftrightarrow\left[2L\right]^{3/2}\sqrt{M_{n}M_{m}E_{l}}H_{mn}=
-g\delta\left(l+P\right)\mathcal{M}\leftrightarrow
-g\frac{L}{2\pi}\delta_{l,-P}\mathcal{M}\] where, from now on, we
shall denote by $H_{mn}$ the matrix element in discrete case,
which then is,\[
H_{mn}=-\frac{cg}{\sqrt{L}}\mathcal{M}\delta_{l,-P}\] where
$c\propto1/\sqrt{M_{n}M_{m}E_{l}}$ . We now first suppose that the
coupling $g$ is weak. Then, we can legitimately employ the first
order time-dependent perturbation theory. Initial state of the
system is the $nth$ bound state. The amplitude that the system is
found in a bound state $m$+ a scalar particle is given by
\cite{MB},\[ C_{m}\left(t\right)=
-i\int_{0}^{t}dt'e^{i\omega_{mn}t'}H_{mn}=
i\frac{cg}{\sqrt{L}}\mathcal{M}\delta_{l,-P}\int_{0}^{t}dt'e^{i\omega_{mn}t'}=
\frac{cg}{\sqrt{L}}\mathcal{M}\delta_{l,-P}\frac{\left(1-e^{i\omega_{mn}t}\right)}{\omega_{mn}}\]
where . So the transition probability is
($\delta_{l,-P}^{2}=\delta_{l,-P}$),
\begin{eqnarray*}
P_{m}\left(t\right) & = & |C_{m}(t)|^{2}\\
 & = & \frac{c^{2}g^{2}}{L}\left|\mathcal{M}\right|^{2}\frac{sin^{2}\left(\frac{\omega_{mn}t}{2}\right)}{\left(\frac{\omega_{mn}}{2}\right)^{2}}\delta_{l,-P}=g^{2}\left|\mathcal{M}\right|^{2}\pi t\frac{c^{2}}{L}\delta\left(\frac{\omega_{mn}}{2}\right)\delta_{l,-P};\qquad t\rightarrow\infty\end{eqnarray*}
where we have employed,\[ lim_{t\rightarrow\infty}\frac{1}{\pi
t}\frac{sin^{2}tx}{x^{2}}=\delta\left(x\right)\] Transition rate, which is probability per
unit time, is \[ R_{m\rightarrow
n}=(const)g^{2}\left|\mathcal{M}\right|^{2}\pi\frac{1}{M_{n}M_{m}E_{l}L}\delta\left(\frac{\omega_{mn}}{2}\right)\delta_{l,-P};\]

\begin{eqnarray*}
\frac{2\pi}{L}\delta\left(\omega_{mn}\right) & = & \frac{2\pi}{L}\delta\left(-M_{n}+M_{m}+\sqrt{l^{2}+m^{2}}+\frac{l^{2}}{2M_{m}}\right)\\
 & = & \frac{2\pi}{L}\delta\left(l-l_{1}\right)\frac{1}{\frac{\partial\omega_{mn}}{\partial l}|_{l=l_{1}}}\leftrightarrow\delta_{l,l_{1}}\frac{1}{\frac{l}{\omega_{l}}+\frac{l}{M_{m}}}\\
 & \simeq & \delta_{l,l_{1}}\qquad assuming\: M_{m}>>l>>m\end{eqnarray*}
Thus, the contribution to the width of $n-th$ state from this decay
$n\rightarrow m$ is \[ \Gamma_{n\rightarrow m}\propto
g^{2}\left|\mathcal{M}\right|^{2}\left[M_{n}M_{m}E_{l}\right]^{-1}\]
The dimensions of $\Gamma$ are that of $M$. Hence, dimensions of
$\left|\mathcal{M}\right|^{2}$ is $M^{2}$. Since the former is a
function of $\xi$ and $\nu$ and $M$ , we have $\Gamma_{n\rightarrow
m}\propto Mf\left[\frac{m}{M}\right]$. Let us apply this to the
transition from 1st excited state to ground state: then, \[
M_{n}=m,\quad M_{m}=m+2.1\left(\frac{M}{m}\right)^{1/3}m,\quad
E_{l}\simeq2.1\left(\frac{M}{m}\right)^{1/3}m,\] so that\[
\left[M_{n}M_{m}E_{l}\right]^{-1}\sim\frac{1}{m^3\times\left(2.1\right)^{2}}\left(\frac{m}{M}\right)^{2/3}\]
This calculation is completed in the Appendix A.
\section{Some Issues}
Before we proceed, we have a number of issues to settle: This will
be done largely in appendices. But, here we list them and state
our conclusions and how we use them. Approximations we made:
\begin{itemize}
\item The constituents of the bound state are non-relativistic.
\item The $O\left(g^{2}\right)$ interactions are sufficient to determine
the bound state structure.
\item Renormalization effects on mass are ignorable. Quantum corrections
to the classical mass of the scalar and whether that can alter our
conclusions.
\end{itemize}
The last issue will be discussed in appendix A. The first two will
be discussed in appendix B.
\section{Identifying Scalar Field with the Stable Bound State}
Consider the full propagator for the scalar field for
$x_{0}>y_{0}$ in presence of bound states. We
have,\begin{eqnarray*}
\Delta_{F}\left(x-y\right) & = & \left\langle 0\right|\phi\left(x\right)\phi\left(y\right)\left|0\right\rangle \equiv\left\langle \Phi\left(x\right)\right.\left|\Phi\left(y\right)\right\rangle \\
 & = & \left\langle \Phi\left(x\right)\right|\mathcal{I}\left|\Phi\left(y\right)\right\rangle \\
 & = & \left\langle \Phi\left(x\right)\right|\left\{ {\displaystyle {\displaystyle \Sigma_{n}\left|\Psi_{n}\right\rangle \left\langle \Psi_{n}\right|}}\right\} \left|\Phi\left(y\right)\right\rangle \end{eqnarray*}
where $\left|\Psi_{n}\right\rangle $is a stable state with the same
quantum numbers as $\left|\Phi\left(y\right)\right\rangle $. As shown
in the earlier section, there is only one such bound state, the ground
state, and there will be a set of scattering states. We shall assume a spectral representation for the exact propagator:\[
\Delta_{F}\left(p\right)\equiv F.T.\left\{ \Delta_{F}\left(x-y\right)\right\} \equiv\intop_{0}^{\infty}\frac{\rho\left(\sigma^{2}\right)}{p^{2}-\sigma^{2}+i\varepsilon}d\sigma^{2}\]
with, \[
\rho\left(\sigma^{2}\right)=Z\delta\left(\sigma^{2}-m^{2}\right)+\mathcal{B}\delta\left(\sigma^{2}-m_{1}^{2}\right)+\rho_{1}\left(\sigma^{2}\right)\]
where, $m_{1}$ is the rest-mass of the ground bound state, $\mathcal{B}>0$,
and we assume   \begin{eqnarray*}
\rho_{1}\left(\sigma^{2}\right) & \geq & 0\\
\rho_{1}\left(\sigma^{2}\right) & = & 0,\qquad\sigma^{2}<4M^{2}\\
\int_{4M^{2}}^{\infty}
\rho_{1}\left(\sigma^{2}\right)d\sigma^{2}&=&1-Z-\mathcal{B}\\&< &1\end{eqnarray*} Here, we have assumed a relation similar to one that is thought to hold in LSZ formulation: $$1=Z+\int^{\infty}_{4M^2}\rho(\sigma^2)d\sigma^2$$. Then, \[
\Delta_{F}\left(p\right)=\frac{Z}{p^{2}-m^{2}+i\varepsilon}+\frac{\mathcal{B}}{p^{2}-m_{1}^{2}+i\varepsilon}+\int_{4M^{2}}^{\infty}d\sigma^{2}\frac{\rho_{1}\left(\sigma^{2}\right)}{p^{2}-\sigma^{2}+i\varepsilon}\]
We note that usually in a QFT, $Z\rightarrow 0$ as more and more channels in propagator are taken into account. Hence, we should have,   $\mathcal{B}+\int_{4M^2}^{\infty}\rho_{1}\left(\sigma^{2}\right)$=1.
Now, we study the propagator for $p^{2}$ small, \[
-\int_{4M^{2}}^{\infty}d\sigma^{2}\frac{\rho_{1}\left(\sigma^{2}\right)}{p^{2}-\sigma^{2}+i\varepsilon}\simeq\int_{4M^{2}}^{\infty}d\sigma^{2}\frac{\rho_{1}\left(\sigma^{2}\right)}{\sigma^{2}}<\frac{1}{4M^{2}}\int_{4M^{2}}^{\infty}d\sigma^{2}{\rho_{1}\left(\sigma^{2}\right)}\lesssim \frac{1}{4M^{2}}\rightarrow 0\]
as $M\rightarrow \infty$, where use has been made of $\int_{4M^{2}}^{\infty}\rho_{1}\left(\sigma^{2}\right)d\sigma^{2}<1$ .\\

[An alternate argument can also be given: The last term in the above depends on $\rho_1\left(\sigma^2\right)$ which in turn depends on probability of finding a scattering state of an invariant mass $\sigma$, where $\sigma>2M$. We expect this to be rather less sensitive to $m$. Now, $C=C\left(g,m,M\right)=C\left[ \frac{g^2}{mM},\frac{m}{M},M\right]$. Now, $C$ is dimensionless, hence $C=C\left[ \frac{g^2}{mM},\frac{m}{M}\right]$.To the lowest order, this quantity is $O(g^2)$ and also insensitive to $m$. Hence, $C \sim \frac{g^2}{mM}\frac{m}{M}\sim \frac{g^2}{M^2}$ .For, similar reasons, we expect that, in higher orders, $C$ is a function of single dimensionless variable $g^2/M^2$ : $C=C\left[g^2/M^2\right]$. In the limit under consideration, $C\rightarrow C[0]=\lim_{g\rightarrow 0} C\left[g^2/M^2\right]=\mbox{the }O(g^2)\mbox{ result}\sim \frac{g^2}{M^2}$. Hence, we have,
\begin{equation}\label{eq:1}
lim_{M\rightarrow\infty}\frac{C}{4M^{2}}=0
\end{equation}]
 and we find that, the corrections to propagator is saturated by the ground bound state. Now, if $Z\rightarrow 0;\mathcal{B}\rightarrow 1$; the propagator itself will be represented fully by the propagation of bound state.

\section{Uses of this Formulation}
\begin{itemize}
\item This formulation allows one to look the scalar field as  representing a propagating bound state.
\item In particular, the interaction Lagrangian of the scalar fields, obtained
by integrating with respect to $\psi$, is that due to a tight bound
state of $\psi,\overline{\psi}$. Non-locality in the interaction
of scalar field can be conceivably understood as due to the bound
state nature of a scalar $\phi$.
\item One cannot distinguish if it is a theory of an elementary field $\phi$
or a composite bound state.
\item This formulation has been useful in study of whether bound state formation can affect causality of the theory\cite{C}.
\end{itemize}

ACKNOLEDGEMENT\\
Part of the work was done when SDJ was "Poonam and Prabhu Goel Chair Professor" at IIT Kanpur.
AH would like to thank NISER for support where part of the work was done.\\
\\
\section*{Appendix A}
In this appendix, we shall perform an explicit calculation of decay width for the process  $\mbox{1st excited state} \rightarrow  \mbox{ ground state + a scalar} $. We have seen, in section 5, that  this is given by
$$\Gamma \propto g^2\frac{1}{(2.1)^2m^3}\left({\frac{m}{M}}\right)^{2/3}\left|\mathcal{M}\right|^2$$
Here, we need estimate $\left|\mathcal{M}\right|^2$ . To do this, Consider first $\mathcal{M}_1$:
$$\mathcal{M}_1 \propto \int_{}^{} {dq}\bar u\left( {q + \frac{P}{2}} \right)u\left( {q - \frac{P}{2}} \right) {\beta _n}\left( {q + \frac{P}{2}} \right)\beta _m^*\left( q \right){E_{q - \frac{P}{2}}} $$
 Now, consider,
\begin{eqnarray*}
\beta _{n}(q)& =& \int_{ - \infty }^\infty dx\;CJ_{\nu }\left( {\xi \exp \left( { - m|x|} \right)} \right)\exp \left( { - iqx} \right) \\
 &=&\frac{1}{m}\int_0^\infty  {dX} C\;J_{\nu }\left( {\xi \exp \left( -X\right)} \right)\exp \left( { - i(q/m)X} \right) + .... \\
 &=& \frac{1}{m}\;C\;F\left[ {\frac{q}{m},\xi } \right] + ....
\end{eqnarray*}
And
\begin{eqnarray*}
C  &=& {\left\{ {\int_{ - \infty }^\infty  {dxJ_\nu ^2\left( {\xi \exp \left( { - m|x|} \right)} \right)} } \right\}^{ - 1/2}} \\
&=& {\left\{ { - \frac{1}{m}\int_\xi ^0 {\frac{{dt}}{t}J_\nu ^2\left( t \right)}  + .......} \right\}^{ - 1/2}}\\
&=&{m^{1/2}}f\left( \xi  \right) \\
\end{eqnarray*}
Thus, we find,
$${\beta _n}\left( q \right) = \frac{1}{{\sqrt m }}\tilde F\left[ {\left[ {\frac{q}{m},\xi } \right]} \right]$$
Setting  ${\frac{q}{m}} = Q$ and letting $m\rightarrow 0 $, we obtain,
 \begin{eqnarray*}
  &=& \int _{}^{} {dq\bar u\left( {q + \frac{P}{2}} \right)u\left( {q - \frac{P}{2}} \right)} {\beta _n}\left( {q + \frac{P}{2}} \right)\beta _m^*\left( q \right){E_{q - \frac{P}{2}}}\\
  &=&  \int_{}^{} {dQ\bar u\left( {mQ + \frac{P}{2}} \right)} u\left( {mQ - \frac{P}{2}} \right){{\tilde F}_n}\left( {Q + \frac{P}{{2m}},\xi } \right)F_m^*\left( {Q,\xi } \right){E_{mQ - \frac{P}{2}}} \\
 \end{eqnarray*}
As $m \rightarrow 0$ and hence $M\rightarrow \infty$, we can set
$$\bar u\left( {mQ + \frac{P}{2}} \right)u\left( {mQ - \frac{P}{2}} \right) \approx \bar u \left( {\frac{P}{2}} \right)u\left( { - \frac{P}{2}} \right)\sim M $$
and also,
$$E_{mQ - \frac{P}{2}} \rightarrow M$$
  Then the expression for $\mathcal{M}_1$  is proportional to
  $$   {M^2}\int_{}^{} {dQ} {{\tilde F}_n}\left( {Q + \frac{P}{{2m}},\xi } \right)F_m^*\left( {Q,\xi } \right){|_{\xi  \to \infty }} $$
  We recognize the above manipulations equivalent to proving:
\begin{eqnarray*}
 \mathcal{M}_1 &\propto& \int_{}^{} {dq}\bar u\left( {q + \frac{P}{2}} \right)u\left( {q - \frac{P}{2}} \right) {\beta _n}\left( {q + \frac{P}{2}} \right)\beta _m^*\left( q \right){E_{q - \frac{P}{2}}}  \\
   &=& M^2 \int_{}^{} {dq} {\beta _n}\left( {q + \frac{P}{2}} \right)\beta _m^*\left( q \right)
  \end{eqnarray*}
  Now,
\begin{eqnarray*}
   &=&\int_{-\infty}^{} {dq} {\beta _n}\left( {q + \frac{P}{2}} \right)\beta _m^*\left( q \right) \\
   &=& \int_{-\infty}^{\infty} {dq} \int_{ - \infty }^\infty  {dxC{J_\nu }\left( {\xi \exp \left( { - m|x|} \right)} \right)} \exp \left( { - i\left[ {q + \frac{P}{2}} \right]x} \right)\\ &\times & \int_{ - \infty }^\infty  {dx'C'{J_\mu }\left( {\xi \exp \left( { - m|x'|} \right)} \right)} \exp \left( {iqx'} \right) \\
  & \propto & \int_{ - \infty }^\infty  {dx} CC'{J_\nu }\left( {\xi \exp \left( { - m|x|} \right)} \right){J_\mu }\left( {\xi \exp \left( { - m|x|} \right)} \right)\exp \left( { - i\frac{P}{2}x} \right) \\
& &\mbox{We employ}\ln \xi  - m|x| = \ln t; \;\;\;dx=\frac{1}{\mp m}\frac{dt}{t} \\
  &=& \frac{1}{m}\int_0^\xi  {\frac{{dt}}{t}CC'} {J_\nu }\left( t \right){J_\mu }\left( t \right)\left[\exp \left( { - i\frac{P}{{2m}}\left( {\ln \xi  - \ln t} \right)} \right)+c.c.\right] \\
  &= &\frac{1}{m}\exp \left( { - i\frac{P}{{2m}}\ln \xi } \right)\int_0^\infty  {\frac{{dt}}{t}CC'} {J_\nu }\left( t \right){J_\mu }\left( t \right)\exp \left( {i\frac{P}{{2m}}\ln t} \right)+c.c.
\end{eqnarray*}
We write,
\begin{equation}
 I = \int_{0}^{\xi}  {\frac{{dt}}{t}} {J_\nu }\left( t \right){J_\mu }\left( t \right)\exp \left( {\frac{{iP\ln t}}{{2m}}} \right)
\end{equation}
We shall write
\begin{eqnarray*}
I &=&  \int_0^{\alpha\sqrt{\nu}}  {\frac{{dt}}{t}} {J_\nu }\left( t \right){J_\mu }\left( t \right)\exp \left( {\frac{{iP\ln t}}{{2m}}} \right)\\ & +& \int_{\alpha \sqrt{\nu}} ^{\xi } {\frac{{dt}}{t}} {J_\nu }\left( t \right){J_\mu }\left( t \right)\exp \left( {\frac{{iP\ln t}}{{2m}}} \right) \\
& \equiv & I_1+I_2
\end{eqnarray*}
where $\alpha$ has been chosen so that the following approximation in the first integral $I_1$  can be made small with any desired degree of accuracy\cite{W}.
$$ J_{\nu}(t) \simeq \frac{t^{\nu}}{\nu !}\qquad ; 0 \leq  t <\alpha \sqrt{\nu}$$
So that $I_1 $ can be ignored

In the second integral, we note: $|J_{\nu}(t)|<1; \mbox{t  real} $. Then\\

$$    |I_2|< \int_{\alpha \sqrt{\nu}} ^{\xi } {\frac{{dt}}{t}}= \ln\frac{\xi}{\alpha \sqrt{\nu}}\sim1/2\ln{\xi} $$
i.e., in fact, not a power-law-behaved quantity.  In fact, we can estimate $I_2$, for
the present transition, as follows: We look at the graphs of $J_{\nu}$ and
$J_{\nu'}$, we find that the product $J_{\nu}J_{\nu'}$ is sharply peaked around a
value $0<<t_0<\xi$, so that to a leading approximation
$$I_2\thickapprox\exp{(iz_0)}\int_{\alpha \sqrt{\nu}} ^{\xi } {\frac{{dt}}{t}} {J_\nu }\left( t \right){J_\mu }\left( t \right)$$
where $z_0=\frac{P\ln{t_0}}{2m}$ Putting this in the expression for $I$ in
$\mathcal{M}_{1}$, we find, using $z=\frac{P\ln{\xi}}{2m}$
\begin{eqnarray*}
\mathcal{M}_{1}&\propto& \frac{M^2}{m}\left[\exp {i(z-z_0)}I_0+c.c.\right]\\
&=& \frac{M^2}{m}\left[2\cos(z-z_0)I_0\right]
\end{eqnarray*}
Noting that
\begin{eqnarray*}
z-z_0&=&\frac{P}{2m}\left(\ln{\xi}-\ln{t_0})\right)
\end{eqnarray*}
\\
Thus, the magnitude of a typical contribution to $\Gamma$ indeed blows in the limit $M \rightarrow \infty$. This implies that the first excited state is totally unstable in this limit. We can see that a similar conclusion plausible for higher excited states.

\section*{Appendix B:The Quantum Corrections to the Mass of the Scalar }
The mass parameter we have been using in the above is the
classical mass. We need to see if the quantum corrections to the
mass of the scalar can alter the conclusions. The quantum
corrections, to $O\left(g^{2}\right)$, can be calculated by
calculating the self-energy of the scalar, now in a model that
admits bound states. As shown in the section 5, in the limit
$g\rightarrow\infty,\: M\rightarrow\infty$, there is only stable
bound state. As argued in section 7, we can
find  the self-energy by effectively saturating the propagator by
the lowest bound state: i.e.we shall estimate the self-energy as
follows: Lowest bound state has a wave-function given by
(\ref{eq:WF}). Let us first consider the scalar on mass-shell with
momentum $p$ with $p^{2}$ on-shell. We can always go to the
rest-frame of the scalar. When the scalar decomposes between a
quark-antiquark pair, they carry momenta
$\frac{q}{2},-\frac{q}{2}+p$ with an amplitude
$\Phi_{0}\left(q\right)$. $\Phi_{0}\left(q\right)$ is the Fourier
transform of the normalized ground state wave-function
$CJ_{\nu}\left(\xi\exp\left[-m|x|/2\right]\right)$.We assume that
the pair is near mass-shell and hence have $q^{2}\approx0$. The
contribution of this intermediate state to the self-energy of
scalar is \begin{eqnarray*} i\Sigma(p) & = & -g^{2}\int
d^{2}k\frac{1}{(2\pi)^{2}}\frac{Tr\left[(\not{k}+M)(\not{k}+\not{p}+M)\right]}{\left(k^{2}-M^{2}\right)\left((k+p)^{2}-M^{2}\right)}\Phi_{0}\left(k\right)\Phi_{0}\left(-k\right)\end{eqnarray*}
In $\Phi_{0}\left(k\right)$, $k$ is effectively a one-momentum. We
can carry out a Wick rotation, so
that\begin{eqnarray*} i\Sigma(p) & = & -g^{2}i\int
d^{2}k\frac{1}{(2\pi)^{2}}\frac{Tr\left[(\not{k}+M)(\not{k}+\not{p}+M)\right]}{\left(k^{2}+M^{2}\right)\left((k+p)^{2}+M^{2}\right)}\Phi_{0}\left(k\right)\Phi_{0}\left(-k\right)\end{eqnarray*}
$k$ now is a Euclidean momentum. We now recall that
$\Phi_{0}\left(k\right)$ falls of rapidly beyond
$\left(k\right)^{2}>\Lambda^{2}$. We can then
estimate the integral by putting a cut-off
$\Lambda$ and find: \[
\Sigma(p)=\frac{g^{2}}{4\pi}\int_{0}^{1}d\alpha\left[\ln\left(\frac{\Lambda^{2}+M^{2}-\alpha(1-\alpha)p^{2}}{M^{2}-\alpha(1-\alpha)p^{2}}\right)-\frac{2\Lambda^{2}}{\Lambda^{2}+M^{2}-\alpha(1-\alpha)p^{2}}\right]\]
For $p^{2}=0$, we have%
\footnote{Mass correction is usually evaluated at $p^{2}=m^{2}$; however the
difference is small and doesn't affect the conclusion.%
}, \[
\Sigma(0)=\frac{g^{2}}{4\pi}\left[\ln\left(\frac{\Lambda^{2}+M^{2}}{M^{2}}\right)-\frac{2\Lambda^{2}}{\Lambda^{2}+M^{2}}\right]\]
Taking into account the momentum behavior of (\ref{eq:WF}), we shall assume
that $\Lambda$ can be chosen as $\sim M^{5/9}$. Then, we can assume
$\Lambda<<M$ and, we have the quantum correction to $m^{2}$, viz.
\begin{eqnarray} \delta m^{2}=-\Sigma(0) & = &
\frac{g^{2}}{4\pi}\left[\frac{\Lambda^{2}}{M^{2}}\right]\end{eqnarray} which goes to
zero as $M^{-2/9}$ as $M\rightarrow\infty$. Thus, the mass of the scalar is stable
against quantum corrections in this limit.\\
\section*{Appendix C: The Bound States and Use of the
Schrodinger Equation}
In using non-relativistic equation, which is
a second order in $g$ i.e. $O\left(g^{2}\right)$, to obtain the
bound states, we have made several approximations:
\begin{itemize}
\item The constituents of the bound state are non-relativistic.
\item The $O\left(g^{2}\right)$ interactions are sufficient to determine
the bound state structure.
\item Renormalization effects on mass are ignorable.
\end{itemize}
We have already addressed to the last question. Further, in our
present case, we have shown that $\frac{<KE>}{2M}\sim\nu^{-2/3}$
(Please see (\ref{eq:keM})). As long as $\nu$ is large, i.e. for
the low lying states, kinetic energy is small compared to the mass
of fermions, and the constituents are non-relativistic. As to the
higher order corrections, we can employ Bethe-Salpeter approach
\cite{BS}. We consider the next order correction to the
non-relativistic momentum space wave-function. Let
$\Phi\left(p\right)$ be the momentum space wave-function as
calculated from the Schrodinger equation. The next order diagram
(See Fig-1) is

\begin{figure}[!htbp]
\centering
\includegraphics[scale=0.4]{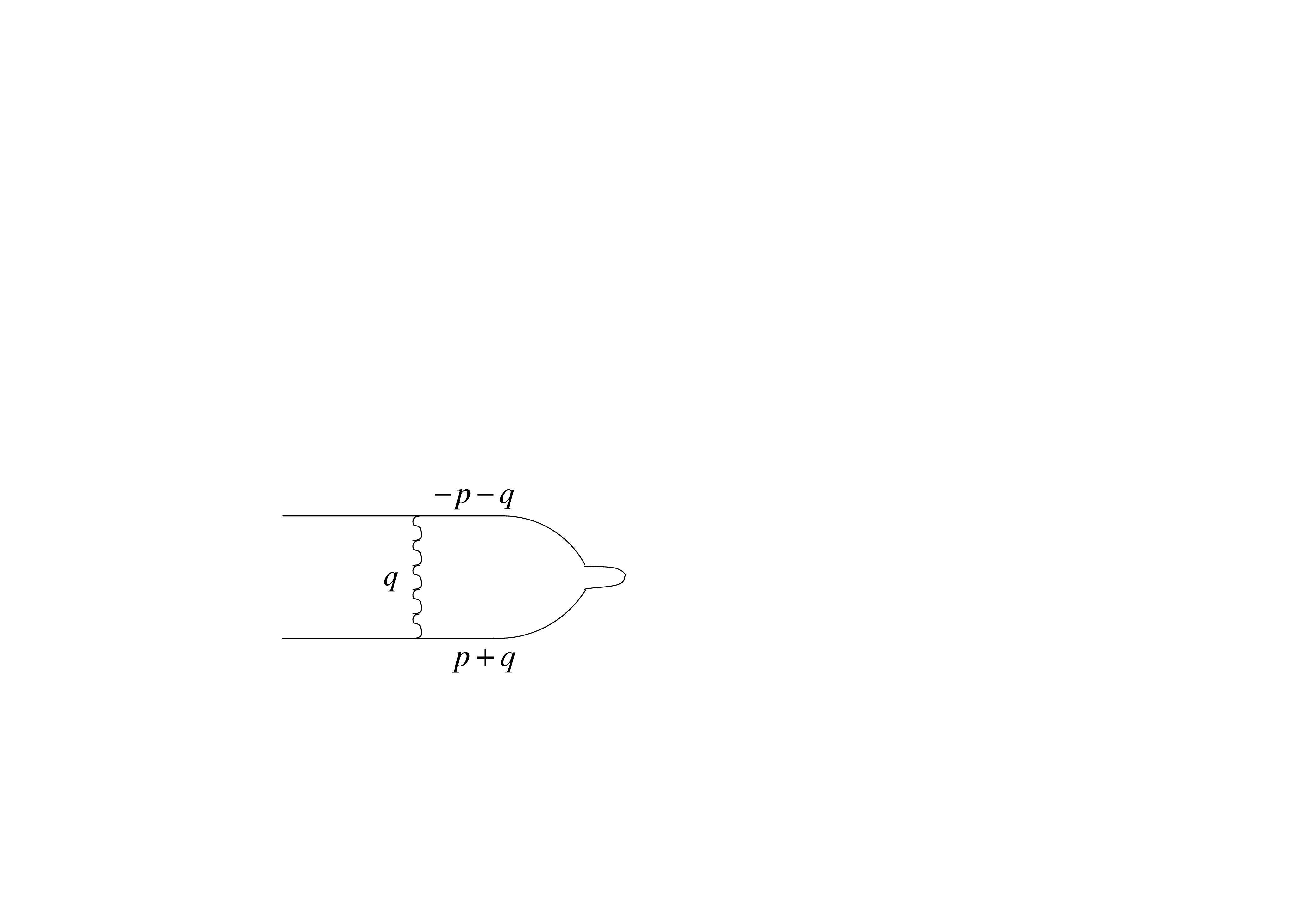}
  \caption{$O(g^4)$ diagram.}
  \label{fig:jI}
\end{figure}
\begin{eqnarray*}
\widetilde{\Phi}(p) & =g^{2} &
\int\frac{d^{2}q}{\left(2\pi\right)^{2}}\frac{i}{p+q-M+i\varepsilon}\Phi\left(p+q\right)\frac{i}{p+q-M+i\varepsilon}\frac{i}{q^{2}-m^{2}+i\varepsilon}\end{eqnarray*}
For the ground state, $\Phi\left(p+q\right)$ is damped out for
$|p+q|\geq O(M^{5/9})$. Thus, the effective range of |p+q| inside
the integral <\textcompwordmark{}< M. It is not difficult to see
that this  integral is suppressed by a dimensionless factor of
$O\left(\frac{g^{2}}{M^{2}}\right)$. [ The singular dependence on $m$  is at worst logarithmic,] Now,\begin{eqnarray*}
\frac{g^{2}}{M^{2}} & = &
\frac{g^{2}}{mM}\frac{m}{M}\sim\frac{m}{M}\ll1\end{eqnarray*}
Thus, in this particular limit, the higher order quantum
corrections are indeed negligible. We note in passing that Harindranath and Perry have
dealt with a problem of bound states between two different species
of fermions \cite{PH} in light-front field theory for the 1+1
dimension Yukawa problem. They have shown the connection between
the $O\left(g^{2}\right)$ quantum correction term and the
Schrodinger equation (Please see Appendix C of reference
\cite{PH}). It holds under the conditions that (i) the
constituents are non-relativistic and (ii) the
$O\left(g^{4}\right)$ and higher order terms in the relevant
equations can be ignored.


\begin{thebibliography}{13}
\bibitem{BC} We have made use of the interactive web-site:
cose.math.bas.bg/webMathematica/webComputing/BesselZeros.jsp
\end{thebibliography}
\end{document}